\documentclass[twocolumn,showpacs,preprintnumbers,amsmath,amssymb]{revtex4}

\usepackage{graphicx}
\usepackage{epstopdf} 
\usepackage{dcolumn}
\usepackage{bm}

\begin{document}


\title{Transition metal impurity-induced generation of plasmonic collective modes in small gold clusters}

\author{N.~Nayyar$^{a}$}
\altaffiliation{Electronic address: nehanayyar@knights.ucf.edu}
\author{A.~Kabir$^{a}$, V.~Turkowski$^{a,b}$, T.S.~Rahman$^{a,b}$}

\affiliation{$^{a}$Department of Physics, 
University of Central Florida, Orlando, Florida 32816\\
$^{b}$NanoScience Technology Center, 
University of Central Florida, Orlando, Florida 32816
}%

\date{\today}

\begin{abstract}
We study the optical properties of small gold chains doped with different transition metal (TM) atoms (Ni,Rh,Fe) by using the time-dependent density-functional theory (TDDFT) approach. The optical absorption spectrum of such systems
demonstrates a collective plasmon mode when the number of atoms is larger than approximately 10, and this mode
splits into two peaks when the Au chain is doped with some of  the TM  (Ni,Fe) atoms. 
We associate the additional peak with a local plasmonic mode
which corresponds to the charge oscillations around the potential created by the d-orbitals of the impurity atoms. The effect takes place when the number of TM atoms is much smaller than the number of Au atoms.
This behavior is opposite to the case of larger bulk noble-metal-TM clusters (radius $>$1nm),
where the doping with TM atoms does not lead to any generation of additional modes, and often leads to a suppression of the main plasmon peak.
This phenomenon of tuning the optical properties of nanosystems with transition metal atoms can be used in many practical applications in the field of nanotechnology.

\end{abstract}



\maketitle

{\it Introduction}.--The physical properties of  bimetallic noble metal-transition metal (TM) nanostructures is an active area of both experimental and theoretical studies, due to a great potential for such systems to be used in modern technologies,
including  catalysis\cite{Bell}, biomedicine \cite{Hirsch}, photophysics \cite{Hodak}, and photonics \cite{Ozbay}
(see also Refs.~\cite{Alonso,Ferrando} for a general review). One of the main reasons for this is that many noble metal systems, like 
bulk and layered extended Au and Ag, have important properties in visible light frequncy range including plasmon excitations, and these properties can be tuned by changing both the geometry
and chemical composition of the clusters, by doping them with TM atoms. Since the TM atoms have both extended s- and
localized d-states, one may expect, similar to the bulk and layered TM systems, interesing and unusual interplay of the role of these two types of states in determining the properties of the systems,
including the optical response.

In the case of noble-metal-TM systems, the structures often have core-shell type, with the TM core. 
For such large particles, the classical Mie's theory can used to study the absorption spectrum\cite{Mie} .
As calculations show, usually the magnitude of the surface plasmon noble metal
peak gets suppressed when the number of TM atoms in the system increases (see, e.g., Refs.~\cite{Alonso,Ferrando}).

Recently, an interest in studying the optical properties of small noble atom chains has arisen, in particular due to a possibility to assemble the Au atom chains on NiAl(110) surface within the STM\cite{Nilius}.
Since the optical properties of such chains, similar to other nanoparticles, would have many practical applications,
the question whether such small systems can demostrate a collective optical response
naturally arises in this case. In general, the Mie's theory is not applicable when the chains are rather
short, therefore one needs to use quantum approaches, like TDDFT. As it was shown by 
Kummel et al\cite{Kummel}, who applied both TDDFT and quantum fluid dynamics in local current approximation (LCA) to small Na atom clusters, one can identify the center-of-mass oscilaltions of the electronic charge
with respect to positive background as plasmons even in the case of rather small clusters, similar to the bulk case. 
Several studies of plasmon modes in noble and alkali metal chains within TDDFT approach are reported in literature. In particular,
Lian et. al.\cite{Lian} studied the possibility of plasmon excitations in Au chains of different length and found that the collective plasmon mode in the adsorption spectrum arises when the number
of atoms is larger than approximately 10. The possibility of central and end plasmon peaks in the case of Na chains with an Ag edge atom was shown by Yan et al\cite{Yan}.They also show that in case of Ag atom chains d electrons can lead to decrease intensity and energy for transverse mode but does not effect much the longitudinal modes\cite{Gao}.Another interesting feature is observed at surfaces by Pohl et.al.\cite{Pohl} where they see acoustic plasmons due to collective effects of surface and bulk electrons which can be the case for future study of chains on NiAL substrate where Au chain electrons and NiAl bulk electrons can lead to such excitations.

To our knowledge, no attention has been paid to the optical properties of TM-doped noble atom chains. 
In order to get some nontrivial different behavior of the bimetallic noble metal-TM nanosystem,
the number of TM atoms must be rather small. On the other hand, one can expect a pronounceable effect from these
atoms only when the number of host (Au) atoms is not too large, when the contribution to the density of states 
from the TM atoms is not negligible. Thus, weakly-doped TM-atom Au chains would be a good candidate for a system, where
one can expect non-trivial results in the optical properties due to TM atoms.

In this Letter, we apply a TDDFT approach to study the optical properties of 2 to 24-atom Au chains weakly
doped with different TM atoms (Ni, Rh and Fe). We demonstrate that in some cases such a doping leads to an unusual response of the system, including generation of extra plasmon peaks.

{\it Computational details.}--To study the optical
absorption spectrum of the chains, we applied a TDDFT aproach by using the Gaussian 03 code
\cite{Gaussian} with B3PW91 hybrid functional\cite{Becke,Perdew} 
and LanL2DZ  basis set\cite{Hay}. This choice of the potential and the basis set allowed us to reproduce reasonably well
the experimental results for the lowest excited states of the Au dimer  
\cite{Bishea,Klotzbuecher}. 
We have considered the case of $Au_{n-m}X_{m}$ chains with the length up to n=24 atoms, some (in most cases one)
of which
were substituted by $1\leq m\leq 4$ X=Ni,Rh and Fe atoms.
In the case of pure Au chains, our results are in a good agreement
with other TDDFT calculations\cite{Lian}. Our main focus was the chains with the interatomic distance
$d=2.89\AA$, observed experimentally in the case of Au chains on NiAl(110) substrate\cite{Nilius}. However,
the cases with different interatomic distances, including the optimized Au-Au and Au-X distances, obtained
for the Au-Au-Au and Au-X-Au trimer chains, were considered. As it follows from our calculations, the optical absorption spectrum does not depends significantly on the interactomic distances even in the case of chains with non-equidistant
nearest-neighbor atoms.
Since the computational time for a given number of excited states increases dramatically with increasing chain size,
we considered the states below 5eV, in particular in order to study all possible absorption peaks in visible range of frequencies.
This choice allowed as to study the longitudinal plasmon modes
in all cases, though we were not able to consider possible high-energy (ultra-violet) transverse plasmon modes, 
found for example in Au chains \cite{Lian}, which is an interesting problem for further studies.
Once the energy levels $E_{i}$ and the transition dipole moments $\langle i|{\hat D}| j \rangle$ were obtained, we calculated the absorption spectrum by using the standard expression 
\begin{eqnarray}
A(\omega )=\sum_{i\not= j} (E_{i}-E_{j})| \langle i|{\hat D}| j \rangle |^{2}e^{-(E_{i}-E_{j}-\omega)/\Gamma}
\label{spectrum}
\end{eqnarray}
(in arbitrary units),
where $\Gamma$ is the peak broadening. In order to make the figures more transparent, in plotted $A (\omega)$ we neglected the contribution of the dipole moments less than 0.1 in atomic units, which produce many peaks
which are much smaller comparing to the main peaks

{\it Results}.--In the case of Au clusters (bondlength $2.89\AA$), we have found that a new (collective) longitudinal plasmonic mode appears
in the optical absorption spectrum when the number of atoms in the even numbered chains is of order 10 (Fig. 1a).
\begin{figure}[t]
\includegraphics[width=7.5cm]{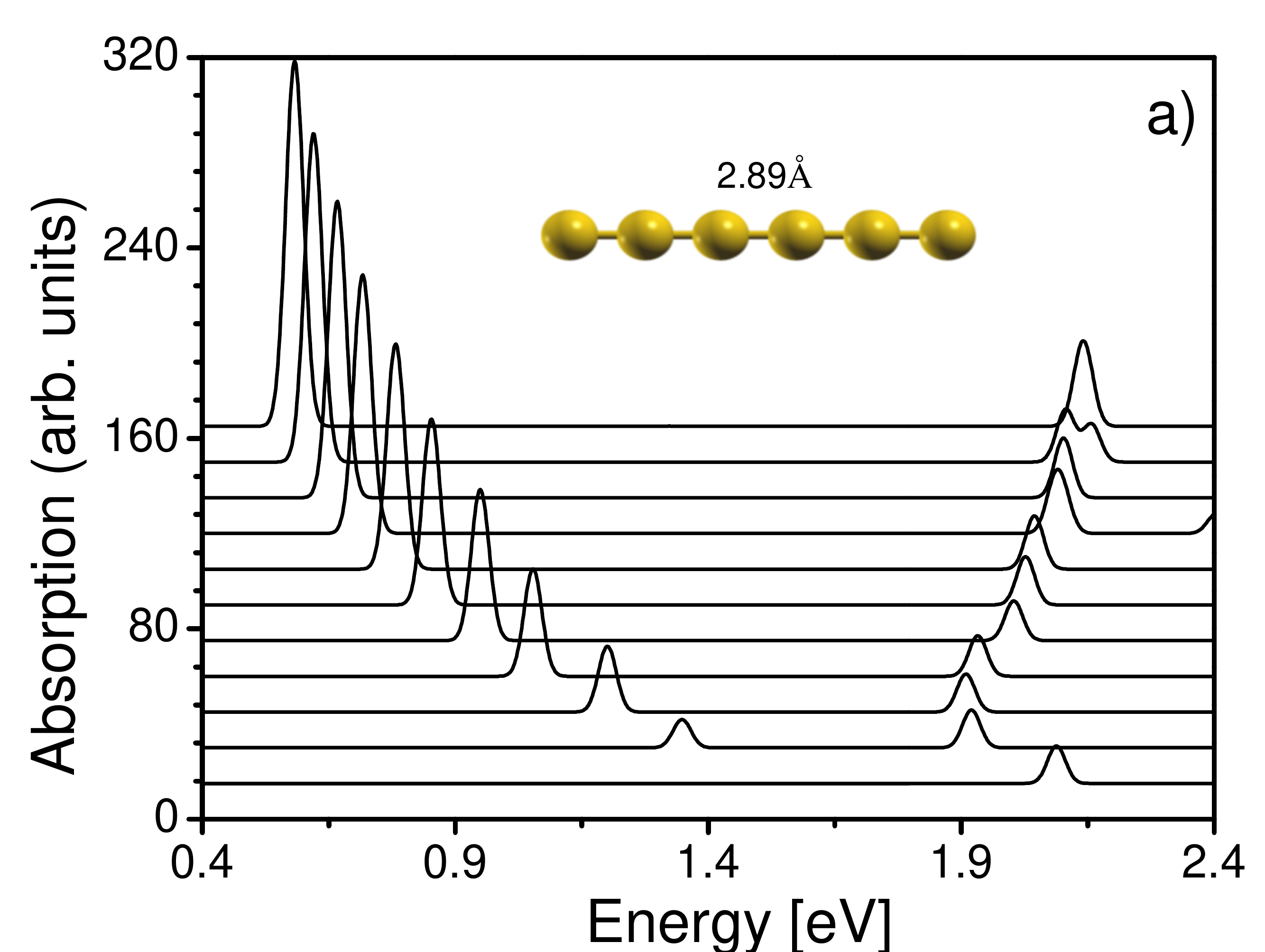}
\includegraphics[width=7.5cm]{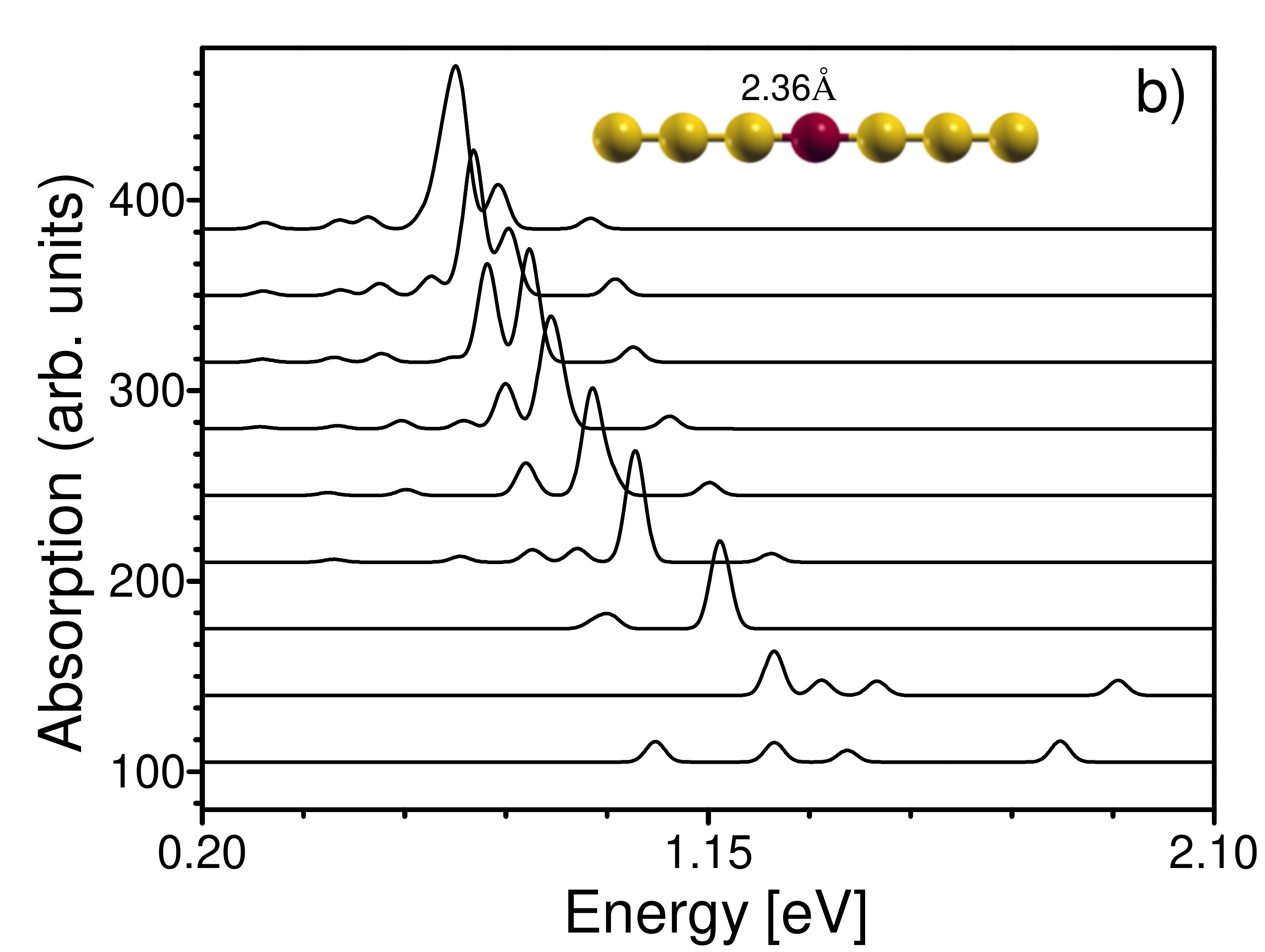}
\caption{\label{fig1} The optical absorption spectrum of the $Au_{n}$ (a) and $Au_{n-1}Ni$ (b) $(2\leq n\leq 24)$ chains. 
Here and in Figures below the interatomic distance is $2.89\AA$. 
}
\end{figure}
The position of the peaks moves to the infrared region with the number of atoms increasing and becomes close to the 
asymptotic value $\sim 0.6eV$ when the number of atoms is around 20. 
Since the magnitude of the peak grows with the number of atom increasing, this shows that
this excitation has a collective form, so with increasing number of atoms more electrons are involved
in the collective motion.
The redshift of the plasmon energy with number of atoms increasing is due to the fact that 
the energy gaps (bandgaps in the many atom case) involved in the dipole excitation are reduced in this case,and hence less energy is required for the transition to excited states 
(see also Ref.~\cite{Lian} the case of Au chains with even number of atoms was also considered).
Interestingly, we have found that in the case of gold chains with odd number of atoms there is a smaller sattelite peak 
with the energy
$\sim 0.1$ below the main peak,
which may be attributed to an additional oscillations related to the unbonded extra charge from the
"odd" atom. Since the study of the difference between the spectra in the case of even and odd number of atoms 
is beyond the scope of this paper, we farther consider only the cases with even number of atoms.
The results for the optical absorption spectrum in the case of Au chains with optimized dimer bondlength
$2.55\AA$ are very similar to Fig.1a, with a small shift of the position of the plasmon peak. 

As the doped case, first we have considered the Au chains with length from 2 to 24 doped with one TM atom X=Ni, Rh and Fe. We have found that in the case of Ni atom there is an additional plasmon peak close in energy
to the main peak (Fig1b). Since this peak does not exists in the case of short chains, it also must have a collective nature.
The facts that this peak appears when one atom is doped and its position almost does not depend on the position
of the Ni atom in the chain (except when it is very close to the chain edge) suggests that it must have a localized nature (see the next Subsection).
We have found a very similar result with a localized plasmon mode in the case of iron atom,
 while there is no such peak in the case of Rh atom.  Since in the Rh atom case the highest occupied s-orbital contains one electron, similar to the Au atom,
while in the Ni and Fe case this orbital is doubly occupied, the presence of the extra peak
may be closely related to the charge redistribution in the doped chains, namely much more stronger involving in the bonding of the d-orbitals in the case of Ni and Fe atoms (see below).
To check this idea we have made calculations also in the case of Ag impurity atom with singly-occupied s-state,
and we have found no extra plasmon peak, similar to the Rh case.

A more close look at the plasmonic modes in the Ni and Fe cases 
when the number of atoms is close to optimal value of long chains with almost no size effect
(chains with the length between 18 and 24 atoms) shows that at such lengths there is change of the plasmon
mode weights when the larger oscillation strength of the dominant higher mode (n=18) moves to the lower one (n=24)
having two (in the case of Ni) or even three (in the case of Fe) close modes with similar strength
when n is close to 20 (see Fig.2, where the results for AuNi chains are presented).
\begin{figure}[t]
\includegraphics[width=7.5cm]{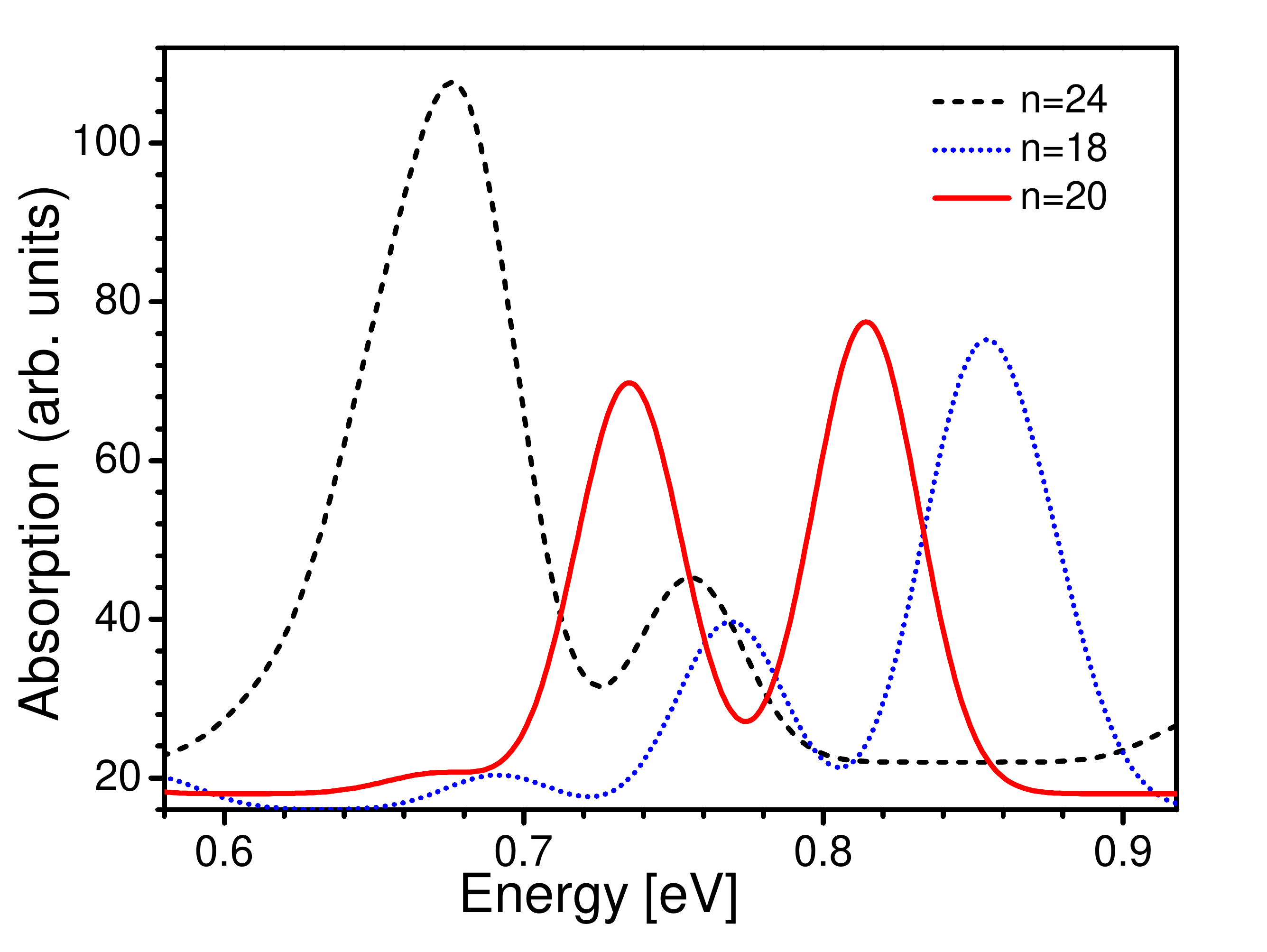}
\caption{\label{fig2} The optical absorption spectrum of the $Au_{n-1}Ni$, n=18, 20, 24, chains.
}
\end{figure}
Since n=20 with the Ni atom in the middle corresponds to two separated chains with the size close to the critical value
$\sim 10$, when the collective extended plasmonic mode appears,
one may assume that at this length there appear strong coherent oscillations of two chains which make the lower mode dominant. The magnitude of the lower peak increases with farther n increasing.
Thus, the second mode can be attributed to the local plasmon, with higher energy energy comparing to the collective
plasmon at $n<20$.The reason for this is that the attractive positive background potential is stronger for the local mode 
(see Figs. 4 and 5 below).
In the case of small number of atoms (between 10 and 20) naturally the collective modes of two subchains (with length
5 to 10) have higher energy comparing to strongly coupled to the background local charge oscillations (see Fig.1a).
The fact that the optical spectrum in the case of long chains almost does not depend on the position of the Ni atom,
except when it is 2-3 atoms close to the edge suggests that the charge distributed over 4-5 Au atoms is involved
in the local oscillations. Fir this reason the position of the local peak depends much less on the chain length comparing
to the collective plasmon, where the energy of the plasmon (spectrum gap) decreases with the number of atom increasing
(at n less than $\sim 20$) 

We have found that the plasmonic peaks in the optical absorption spectrum disappear when 
the number of the impurity atoms is around 5. The results for the case of quasi-equidistant Ni impurities
far from the edges
in the 24-atom Au chain are presented in Fig.3.
\begin{figure}[t]
\includegraphics[width=7.5cm]{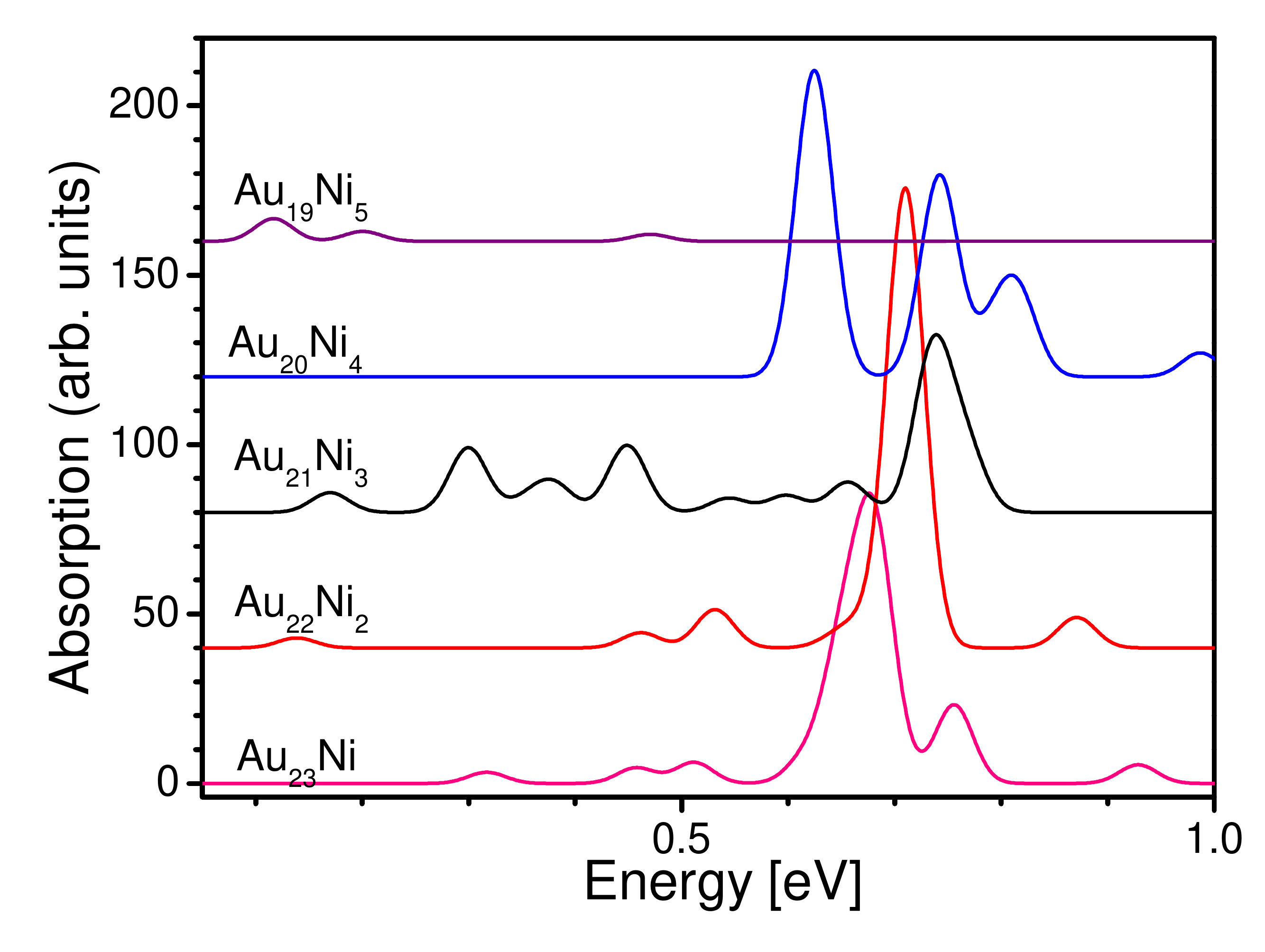}
\caption{\label{fig3} The optical absorption spectrum of the $Au_{24-n}Ni_{n} (n=1,..,5)$ chains.
}
\end{figure}
Indeed, in the case of 5 impurities, or in the case of five 3-atom and one 4-atom separated Au chains,
the chains are too short to excite either collective or local plasmon excitation (see Fig.1).
Similar results were found in the clubbed case.
In particular, in the case of $Au_{18}Ni_{2}$ there are two plasmon peaks in both separated
and clubbed cases, which are separated by distance $\sim 0.1eV$, but the center of the peak positions
is shifted in the clubbed and separated cases.
This result also suggests that there are two kinds of excitations in the system: collective motion in subchains
and some local charge oscillations near the AuNi border.
 
{\it Discussion.}--In order to get a better understanding of possible reasons of the plasmon peak splitting,
we analyze the Mulliken atomic charge distributions in the case of Au and single-doped Au clusters.
The results for the $Au_{20}$, $Au_{19}Ni$ and $Au_{19}Rh$ are presented in Fig.4
\begin{figure}[t]
\includegraphics[width=7.5cm]{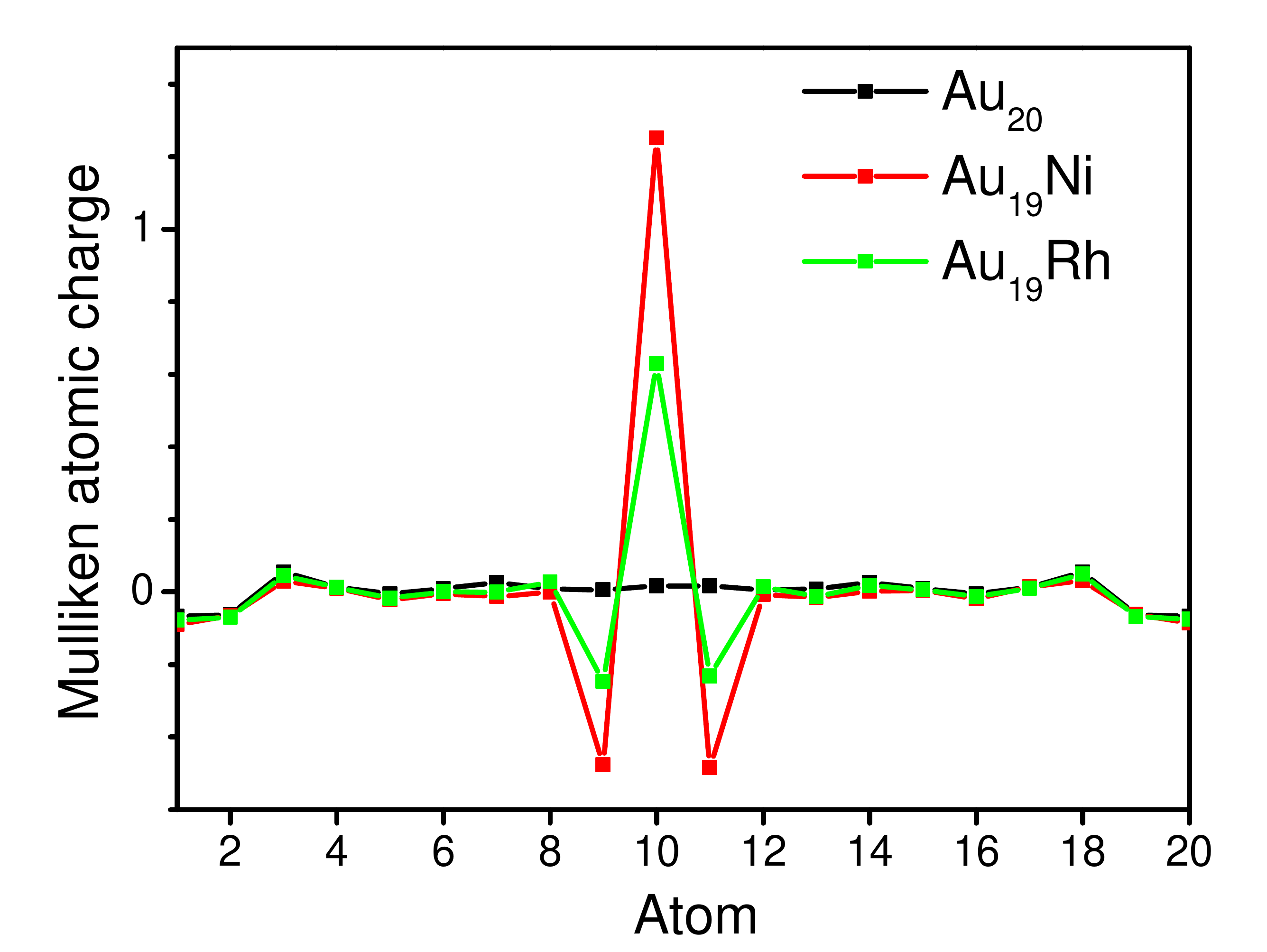}
\caption{\label{fig5} The Mulliken atomic charge distribution in the case of $Au_{19}X (X=Au,Ni,Rh)$. The impurity atom
is put at site 10 from the left.}
\end{figure}
(the results for the $Au_{19}Fe$ case is very similar to $Au_{19}Ni$).
Since the charge distribution on the edges is almost identical in all three cases and there is no second plasmon peak in the 
pure Au chain, it means that the edge oscillations are not responsible for the second peak in the doped chains.
On the other hand, there is very strong difference in charge
distributions near the impurity atom at the center of the chain in all three cases. In the case of Ni atom,
the impurity ion has a very large positive charge, while the nearest Au atoms are negatively charged.
One can model this situation as a jellium electron model in presence of one high potential well
and two neighbor valleys, and the rest of the chain can be considered as the standard positive background
in the plasmon problems (Figure 5).   
\begin{figure}[t]
\includegraphics[width=7.5cm]{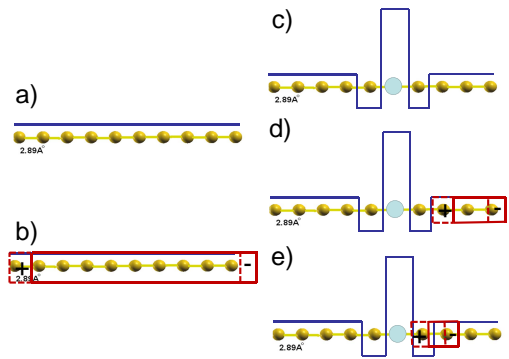}
\caption{\label{fig5} Schematic form of the positive charge background potential for the $Au_{10}$ (a )and $Au_{9}X$ (c) chains.
In the case of $Au_{10}$, the plasmon excitation corresponds to the charge oscillation along all the chain (b).
In the case of $Au_{9}X$ chains, there are possible two kinds of plasmonic oscillations: collective plasmon
excitations in each half of the chain (d) and local plasmon oscillations on both sides of the impurity atom potential wall (e).}
\end{figure}
The plasmon oscillations can be modeled as local mode and a collective mode in the case of atom with impurities
(Fig.5d and 5e). To understand the difference between the results for the Rh and Ni(Fe) atoms, one can
find from Fig.4 that the height of the potential well and the depth of the valleys are much smaller in the Rh case.
In this case, when the charge from the right subchain close to the valley moves to the right but the valley potential
is not deep enough to move the charge back, there is no reason for the local oscillations.
To estimate the necessary potential strength, one can assume that the restoring energy effects
are important when this energy is of order or larger of the kinetic energy in the system.
The kinetic energy of electrons can be estimated as twice nearest-neighbor hopping energy.
We approximate it by the expression for the Slater-Koster parameter $V_{ss\sigma}=1.32\hbar^{2}/md^{2}$ multiplied by 2, where m
is the electron effective mass, which we put equal to the free electron mass $m_{e}$ for simplicity
and $d=2.89\AA$ is the inter-atomic distance.
This gives $E_{kin}=1eV$ (Actually, this energy can be lower since the effective electron mass in the gold chain can be much larger than $m_{e}$). The potential energy is of order of local plasmon frequency $\omega_{lp}$,
which can be estimated from the bulk formula $\omega_{lp}=\sqrt{4\pi ne^{2}/m}$, where n is the electron density.
To find the value of local density, we can use the ratio of the plasmon frequencies in the bulk and in the chain,
which must be equal to the ratio of the square roots of the corresponding densities.
Since in the bulk the density is $5.9\times 10^{22}cm^{-3}$ and the plasmon frequency - 5.26eV, and the chain plasmon
frequency is $\sim 0.66eV$, this gives the chain charge density $7.68\times 10^{20}cm^{-3}$, or 1 electron per atom.
If one models the chain as a tube of some radius and length $N_{atoms}\times 2.89\AA$, this density corresponds
to the tube radius $12\AA$.
In the case of the doped chain, when the impurity charge is shifted to the nearest Au atoms, the local charge density increases,
which leads to an increase of $\omega_{lp}$. From the estimations above, in order to get the local plasmon
oscillations one need to have $\omega_{pl}\sim E_{kin}\sim 1eV$, which corresponds to the following number of
electrons per atom: $(1eV/0.66eV)^{2}\sim 2$. In other words, the nearest to the impurity atoms must get one extra
electron in order to generate local plasmons. As we see from Fig.4, this analysis gives semi-quantitatively
correct results in the case of Ni and Rh atoms. In the first case, the extra charge is $\sim 0.5$ electron, i.e. of order of
one electron, while in the case of Rh atom it it almost order of magnitude smaller.

One can analyze the reason of the charge redistribution from the bonding point of view.
Since there is only one valence s-electron
in the case of Rh atom, this electron participates in $\sigma$
bonding, while the other eight d-electrons are not very chemically active. Thus, one can assume that the charge of the Rh
core is +1, similar to the Au atoms and the positive and negative charges of the $Au_{n-1}Rh$ and  $Au_{n}$ "plasmas"
are almost the same, which results in the same type of the excitations, i.e. one plasmon.
In the case of Ni and Fe atoms, the valence s-state is doubly occupied which means that mainly d-states contribute 
to the atomic bonding.  A more accurate look shows that $d_{z}^{2}$, $d_{xz}$ and $d_{yz}$ orbitals donate most of the charge to the bonding, which results in a positively charged core with charge $Z>+1$.
This is the main reason for the potential well at the impurity atom.
Thus, one can estimate which sorts of atoms may lead to generation of localized plasmon looking
for the bonding charge redistribution. Obviously the estimation presented above is not very accurate,
and more refined estimations can be easily obtained by improving this one, in particular by using
correct electron effective mass. However, we believe that even such simple estimations give correct order of magnitude of the energy of  possible plasmon excitations.
  
To summarize, we have shown that a weak doping of Au chains with some TM atoms may lead to generation of local plasmonic modes. This effect is a result of a delicate balance between the s- and d-states of TM atoms,
which may lead to a strong charge redistribution in the chain near the impurity atom. Namely, the double occupancy
of the top s-orbital and partial occupancy of the d-states are necessary in this case. We have also proposed a simple 
criteria for possibility of the local plasmon generation based on the charge redistribution between the impurity and nearest host atoms. These results show that some TM atoms can be used to tune the optical properties of nanostructures, 
which can have many possible areas of applications: from solar cells to sensors.
Several important question related this problems remain open, including the role of the substrate
in the optical response of the chains and possibility of magneto-plasmons generated 
by ferromagnetically-ordered TM impurities (see,e.g., Ref.\cite{Weick}). 
We are planning to study these problems in nearest future.

\section*{Acknowledgments}
We would like to acknowledge a finantial support from DOE under Grant No. DOE-DE-FG02-07ER46354.


\begin{thebibliography}{99}

\bibitem{Bell}
A.T. Bell, Science {\bf 299}, 1688 (2003)

\bibitem{Hirsch}
L.R.~Hirsch, R.J.~Stafford, J.A.~Bankson et al, 
Proc. Natl. Acad. Sci. U.S.A. {\bf 100}, 13549 (2003). 

\bibitem{Hodak}
J.H.~Hodak, A.~Henglein, and G.V.~Hartland,
J. Phys. Chem. B, {\bf 104}, 9954 (2000).

\bibitem{Ozbay}
E.~Ozbay, Science {\bf 311}, 189 (2006).

\bibitem{Ferrando}
R.~Ferrando, J.~Jellinek, and R.L.~Johnston,
Chem.~Rev. {\bf 108}, 845 (2008).

\bibitem{Alonso}
J.A.~Alonso, {\it Structure and Properties of Atomic Nanoclusters}
(Imperial College Press, London, 2005).

\bibitem{Mie}
 G.~Mie, Ann. Phys {\bf 330}, 3371 (1908).

\bibitem{Nilius}
N.~Nilius, T.M.~Wallis, and H.~Ho, Science {\bf 297}, 1853 (2002).

\bibitem{Kummel}
S. Kummel, K. Andrae and  P.G. Reinhard Appl.Phys. B :Lasers Opt. {\bf 73},293 (2001).

\bibitem{Lian}
K.-Y.~Lian, P.~Salek, M.~Jin, and D.~Ding, J.Chem.Phys.{\bf 130}, 174701 (2009).

\bibitem{Yan}
J.~Yan , Zh.~Yuan and Sh.~Gao, Phys. Rev. Lett. {\bf 98}, 216602 (2007).

\bibitem{Gao}
J.~Yan and Sh.~Gao, Phys. Rev.B{\bf 78}, 235413 (2008)

\bibitem{Pohl}
K.~Pohl , B.~Diaconescu ,G.Vercelli , L.Vattuone , V.M. Silkin , E.V. Chulkov , P.M. Echenique and M.Rocca , EPL,{\bf 90} (2010) 57006

\bibitem{Gaussian}
M. J. Frisch and G. W. Trucks, and H. B. Schelgel et al., {\it GAUSSIAN 03, revision D.01}, 
(Gaussian, Inc., Wallingford, CT, 2004).

\bibitem{Becke}
A.D.~Becke,  J. Chem. Phys., {\bf 98}, 5648 (1993).

\bibitem{Perdew}
J.P.~Perdew and Y.~Wang, Phys. Rev. B {\bf 33}, 8800 (1986);
J.P.~Perdew, {\it Electronic Structure of Solids},  Eds. P. Ziesche and H Eschrig (Berlin: Academic, 1991).

\bibitem{Hay}
P. J. Hay and W. R. Wadt, J. Chem. Phys. {\bf 82}, 299 (1985).

\bibitem{Bishea}
G.A. Bishea and M. D. Morse, J. Chem. Phys. {\bf 95}, 5646 (1991).

\bibitem{Klotzbuecher}
W.E.~Klotzbuecher and G. A.~Ozin, Inorg. Chem. {\bf 19}, 3767 (1980).

\bibitem{Weick}
G.~Weick, D.~Weinmann,
Phys. Rev. B {\bf 83}, 125405 (2011). 

\end{thebibliography}
\end{document}